\documentclass[aip,graphicx,amsmath,amssymb]{revtex4-1}
\usepackage{graphicx}
\usepackage{caption}
\usepackage{subcaption}
\usepackage{xurl}

\newcommand{\bs}[1]{\boldsymbol{#1}}

\draft 

\begin{document}


\title{Effective control of two-dimensional Rayleigh--B\'enard convection: invariant multi-agent reinforcement learning is all you need} 



\author{Colin Vignon}
\affiliation{FLOW, Engineering Mechanics, KTH Royal Institute of Technology, SE-100 44 Stockholm, Sweden}
\affiliation{Mines de Paris, Universite PSL, 75005 Paris, France}

\author{Jean Rabault}
\affiliation{IT Department, Norwegian Meteorological Institute, Postboks 43, 0313 Oslo, Norway}

\author{Joel Vasanth}
\affiliation{FLOW, Engineering Mechanics, KTH Royal Institute of Technology, SE-100 44 Stockholm, Sweden}

\author{Francisco Alcántara-Ávila}
\affiliation{FLOW, Engineering Mechanics, KTH Royal Institute of Technology, SE-100 44 Stockholm, Sweden}

\author{Mikael Mortensen}
\affiliation{Department of Mathematics, University of Oslo, 0316 Oslo, Norway}

\author{Ricardo Vinuesa}
\email[]{rvinuesa@mech.kth.se}
\affiliation{FLOW, Engineering Mechanics, KTH Royal Institute of Technology, SE-100 44 Stockholm, Sweden}


\date{\today}

\begin{abstract}

Rayleigh--B\'enard convection (RBC) is a recurrent phenomenon in a number of industrial and geoscience flows and a well-studied system from a fundamental fluid-mechanics viewpoint. In the present work, we conduct numerical simulations to apply deep reinforcement learning (DRL) for controlling two-dimensional RBC using sensor-based feedback control. We show that effective RBC control can be obtained by leveraging invariant multi-agent reinforcement learning (MARL), which takes advantage of the locality and translational invariance inherent to RBC flows inside wide channels. MARL applied to RBC allows for an increase in the number of control segments without encountering the curse of dimensionality that would result from a naive increase in the DRL action-size dimension. This is made possible by the MARL ability for re-using the knowledge generated in different parts of the RBC domain. MARL is able to discover an advanced control strategy that destabilizes the spontaneous RBC double-cell pattern, changes the topology of RBC by coalescing adjacent convection cells, and actively controls the resulting coalesced cell to bring it to a new stable configuration. This modified flow configuration results in reduced convective heat transfer, which is beneficial in a number of industrial processes. We additionally draw comparisons with a conventional single-agent reinforcement learning (SARL) setup, and report that in the same number of episodes, SARL is not able to learn an effective policy to control the cells. Thus, our work both shows the potential of MARL for controlling large RBC systems, as well as demonstrates the possibility for DRL to discover strategies that move the RBC configuration between different topological configurations, yielding desirable heat-transfer characteristics.
\end{abstract}


\maketitle 

\section{Introduction}\label{introduction}

Deep reinforcement learning (DRL) has recently emerged as a promising methodology for performing active flow control \cite{rabault_kuhnle_2023} (AFC). DRL is a subclass of the much more general field of RL, where deep artificial neural networks are used to parameterise the value function in Q-learning methods \cite{mnih2015} or serve to parameterise the policy or the mapping from state to actions in policy gradient methods \cite{lecun2015deep, sutton2018reinforcement}. The DRL ``controller'', which includes the control law to be applied on the system, and algorithms around it to allow tuning of the control law and query of the control law, is called the DRL agent. The general principle of the (D)RL paradigm \cite{li2017deep} is based on considering the system to control as a black box (the ``plant'' in classical control theory, or the ``environment'' in DRL terms), and to interact with it through just three channels of communication. At a given timestep $t$, i) the DRL agent can query the environment for a (possibly noisy, stochastic) state estimate $s_t$, ii) the DRL agent can provide to the environment a control instruction called the action in DRL terms, $a_t$, iii) the environment can provide back a reward $r_t$ that indicates the quality of the current state of the system. Based on these three channels of interaction, the DRL paradigm aims at defining effective algorithms to learn control strategies that maximize the combined expected rewards at all future times from the given state at time $t$, also known as the  expected return $G_t = \sum_{k=0}^{\infty} r_{t+k+1} \gamma^k$, where $\gamma$ is the discount factor, typically $\gamma \approx 0.95-0.99$. In order to make the computation of $R_t$ tractable, DRL interaction with the environment takes place on an episode basis during training. An episode is a simulation with a given time extent (typically covering a few timescales of the slowest dynamics of the system), and corresponds to one individual trajectory of the system in its phase space. The optimization of the control applied by the DRL agent is performed through trial-and-error learning and direct online interaction with the environment, a process that involves stochastic exploration of the environment properties (referred to as the `training mode' of the DRL agent). This allows to explore and map the behavior of the environment and to determine the best trajectory that the environment should be actuated to follow in the phase space to optimize the reward. A number of algorithms have been developed to perform this optimization, the most famous belonging to either i) the Q-learning paradigm \cite{van2016deep, mnih2013playing} based on the Bellman equation, ii) the policy-gradient method \cite{silver2014deterministic}, which relies on a direct estimate of the sensitivity of the reward to the policy parameters, and iii) a combination of both paradigms, such as the actor-critic proximal-policy optimization (PPO) method \cite{schulman2017proximal}, which combines their respective strengths. Once the trial-and-error learning process has converged and the DRL agent is fully trained, the trial-and-error exploration component can be turned off, so that a deterministic controller is obtained (corresponding to the  `deterministic mode' of the DRL agent).

This DRL framework is very attractive for the control of complex, high-dimensional, non-linear systems such as those obtained in AFC. Indeed, DRL does not make any assumptions on the system to control; actual DRL algorithms have been increasingly refined in the course of the last 10 years and are well adapted to solve non-linear, high-dimensional control problems from games \cite{mnih2013playing, silver2018alphago, vinyals2019grandmaster}, industrial control problems \cite{degrave2022magnetic} or classical physical systems such as those present in fundamental fluid mechanics \cite{sonoda2022reinforcement, guastoni2023deep}. Compared with traditional approaches, such as the adjoint method \cite{chevalier2002adjoint}, DRL does not need direct insight into the full state of the system and its governing equations. Moreover, while DRL training is expensive in terms of the number of interactions with the environment needed for the trial-and-error learning to take place, once the DRL agent is trained, predicting the next action to take given the current state is very cheap computationally. These aspects make DRL an attractive method for real-world control applications, including AFC.

Over the last 5 years, DRL for AFC has been applied to increasingly complex cases. Starting from simple 2D cylinder cases \cite{rabault2019artificial} at low Reynolds number ($Re = UL / \nu$, which measures the relative importance of inertia and viscosity, where $U$ is the characteristic velocity scale of the system, $L$ its characteristic length scale, and $\nu$ the kinematic viscosity), DRL has further been used for investigating a number of different configurations involving 2D cylinders \cite{xu2020active, qin2021application, tokarev2020deep, li2022reinforcement}, looking into increasingly complex stochastic wake control problems as $Re$ is increased \cite{ren2021applying, varela2022deep}, and recently, considering the control of 3D channel flows \cite{sonoda2022reinforcement, guastoni2023deep}. DRL has also been applied to a number of other AFC cases, from 2D Rayleigh--B\'enard convection in a small-size channel \cite{beintema2020controlling}, to instabilities developing in thin fluid films \cite{belus2019exploiting}, engineering cases related to wings \cite{wang2022deep, vinuesa2022flow} and other practical problems \cite{ren2019active, ren2021bluff, jiang2022artificially}. DRL has also been combined with flow-stability analysis to guide the placement of observation probes~\cite{li2022reinforcement} for effective control. In this case, a stability analysis was used to reveal the structural sensitivity that determines the location of the origin of the instability, i.e., the ‘wavemaker’ region. The structural sensitivity can also be computed using data-driven techniques~\cite{corrochano22structural}. Positioning the observation probes in the region of the wavemaker has shown to achieve the best possible learning curve with the reward maximization then taking place in the least number of episodes, compared with other configurations of observation probes~\cite{li2022reinforcement}. While these studies have mostly considered numerical test cases, demonstrations of the applicability of DRL for real-world AFC have been presented in test experiments \cite{fan2020reinforcement}.

Moreover, a number of peripheral challenges associated with discovering efficient flow control strategies have been solved over the last few years. Rabault and Kuhnle~\cite{rabault2019accelerating} demonstrated how the multi-environment paradigm can be leveraged to obtain speedups in training speed proportional to the number of environments used. In particular, they showed that DRL training speed is directly proportional to the number of (state, action, reward) triplets generated per wall clock time unit, and that this is in turn directly proportional to the number of CFD simulations used in parallel to train the DRL algorithm and feed it with data streams. Belus et al. \cite{belus2019exploiting} have demonstrated that the curse of dimensionality on the control dimension can be effectively solved by leveraging a multi-agent reinforcement-learning (MARL) approach (although the authors did not refer to MARL under this name at the time), in cases where some general properties of the system are invariant in space. In particular, MARL was demonstrated to be an effective way to control systems containing several generally similar structures that are far enough to be mostly independent from each other, by sharing the knowledge acquired among them. This, in turn, allows to reduce the combinatorial cost and alleviate the curse of dimensionality associated with the exploration of large action spaces.

Following these advances, DRL can now be used to control different physics as the effect of non-linearity increases \cite{varela2022deep}, to perform effective control over a range of flow conditions in simple AFC test cases \cite{tang2020robust}, or to control complex three-dimensional (3D) turbulent channel flows \cite{guastoni2023deep}. DRL has also been demonstrated in a number of test situations involving chaotic flows, such as the one arising from the one-dimensional (1D) Kuramoto--Sivashinsky system of equations \cite{bucci2019control, xu2023reinforcement}. The placement and optimization of the sensor inputs used to drive DRL algorithms has also been investigated, see Refs.~\onlinecite{paris2021robust, paris2023reinforcement}. As the literature is growing fast, we refer the reader curious of a more detailed overview of DRL applications to fluid mechanics in general and AFC in particular to one of the several reviews that have recently been provided on the topic\cite{rabault2020deep, garnier2021review, doi:10.1063/5.0143913}. Similarly, DRL is becoming an increasingly common topic in AFC, and the DRL methodology \textit{per se} has now been discussed in a number of AFC studies, so we refer the reader curious of specific DRL algorithmic details and refinements to the corresponding body of literature \cite{matsuo2022deep}. As a consequence of this increasing popularity, DRL frameworks are starting to emerge, both in the DRL~\cite{tensorforce, raffin2021stable, weng2021tianshou} and in the fluid-mechanics communities~\cite{wang2022drlinfluids, kurz2022deep, kurz2022relexi}, a fact that now provides frameworks for effectively coupling DRL with well-established computational-fluid-dynamics (CFD) tools. This makes it increasingly easier to apply DRL to AFC problems.

At this point, we want to emphasize that, while DRL is a promising method given that it has provided state-of-the-art results in a variety of fields as previously discussed, it is not the only method that can perform AFC by considering the flow to control as a black box to be optimized. Machine-learning control (MLC, \cite{brunton2015closed, duriez2017machine, noack2019closed}) has, indeed, been a topic of discussion for quite a long time, and the DRL method is, in this aspect, a newcomer in this field. Traditionally, methods such as genetic programming (GP \cite{debien2016closed, li2017drag}), or Bayesian optimization (BO \cite{blanchard2021bayesian}), have also been applied with great success to a variety of practical problems \cite{gautier2015closed, zhou2020artificial, raibaudo2020machine, maceda2021stabilization} and can be competitive with DRL in a number of applications \cite{pino2023comparative}. However, we focus our efforts on DRL in the present study, as DRL is showing great flexibility in constraining the admissible structure of the control law (see, e.g., the discussions around invariants and MARL in Ref.~\onlinecite{belus2019exploiting}). Moreover, the DRL methodology is promising thanks to its ability to use local (in space, but also in time) information to perform optimization. Compared to, e.g., a direct application of GP or BO, which only optimize based on the mean efficiency of a control law over one simulation containing several typical physical time scales of the underlying dynamics, and which do not exploit information about the high-frequency quality of the control law, DRL gets a reward at each interaction step. Thus, DRL can learn both to reproduce positive sequences of actions and avoid negative sequences of actions happening within a single episode. As a consequence, DRL is able to generate information about the quality of the control law at a higher time granularity than the episode itself, and generates a larger volume of usable reward information than would be provided by monitoring for example the averaged reward over an episode.

In the present work, we use DRL techniques to discover effective strategies to control the instabilities and the heat flux in a two-dimensional (2D) Rayleigh--Bénard convection (RBC) problem. The RBC phenomenon is intrinsic to many industrial applications and can be found in a wide range of natural phenomena \cite{get98}. The canonical initial condition in a RBC problem is a fluid at rest that is being heated from a lower wall and/or cooled from an upper wall (see Fig.~\ref{setup_sketch}). Because of the induced temperature gradients and due to the fluid-buoyancy effects, natural convection occurs. The dynamical state of the RBC system can be characterised by a non dimensional parameter, the Rayleigh number $Ra$, which is a ratio of the time scale of heat transport by diffusion or conduction to the time scale of heat transport by buoyancy-induced convection. Beyond a critical value of $Ra=Ra_c$ when convective effects dominate, RB instability occurs through a supercritical bifurcation from an initial quiescent state, to time-dependent cellular motion \cite{jietang94}.

The convective cells cause a non-spatially uniform heat transfer along the domain, as well as large-scale fluid motions and structures containing significant kinetic energy, which can be detrimental in many industrial applications. Thus, control of RBC is crucial to maximize the efficiency and quality in industrial processes or systems in various sectors. The control might involve, for example, modulating the spatial distribution of the bottom-plate heating in the canonical RBC configuration. Optimizing such a control is challenging topic for classical control-theory methods. Indeed, many classical linear-control techniques such as the adjoint method have only a limited domain of applicability, owing to their local, linear nature, which limits their ability to control non-linear phenomena such as RBC.

Passive flow control has been extensively used with minor improvement in the $Ra_c$ for which instabilities start \cite{dav74,don90,kel92,car14,swa18}. On the other hand, AFC has explored more sophisticated control strategies that manage to keep a stable convective cell at considerably higher Rayleigh numbers than $Ra_c$. The most common way of carrying out this AFC has been with the use of jet controllers, performing small perturbations of the velocity and/or temperature, or with a linear control of the temperature in the heated boundary layer \cite{sin91,wan92,tan93a,tan93b,tan94,tan95a,tan95b,sho96,how97a,how97b,tan98a,tan98b,how00,or01,or03,or05,rem07}. In configurations where the RB instability is too strong to be completely eliminated, a reduction in the strength of the RBC cells and associated flow motion, as well as kinetic energy, are also key goals \cite{beintema2020controlling}. Recently, the use of DRL to control the RBC has been employed for the first time in Ref.~\onlinecite{beintema2020controlling}. Comparing the results with the linear control employed in Ref.~\onlinecite{rem07}, the $Ra_c$ beyond which instabilities start to emerge in the specific configuration considered was increased from $10^4$ to $3\times 10^4$, and even in the case where RBC cells were present, their intensity was strongly reduced.

In the present work, we extend the work in Ref.~\onlinecite{beintema2020controlling} (which we will refer to in the rest of this paper as GB for ``Gerben Beintema'', from the name of the lead author), to look further into DRL control of RBC. In particular, we improve on the following aspects: 

\begin{itemize}
    \item The domain used by GB is bounded by adiabatic walls on the right and left ends and the domain itself is relatively narrow. This in-turn constrains the RBC cells obtained. Moreover, the walls presented in GB's setup can be used by the DRL agent to help move around and break the RBC cells, as they provide strong flow confinement. By contrast, we use a wider domain of height $H=2$ and width $L = 2\pi$ (compared with the $H=1$ and $L = 1$ used in GB) with periodic conditions on the right and left boundaries. In other words, our domain has an aspect ratio of $\pi$ compared with the aspect ratio of $1$ used in GB. As a consequence, our DRL agent has to control the RBC instabilities without the possibility to help itself with the walls, and the RBC cells are closer to unconstrained cells obtained in wide domains relevant for industrial use.

    \item One limitation in GB's work is how DRL control is applied to the RBC simulation. In order to control the RBC cells, GB modulates the temperature at the lower wall, keeping its mean temperature constant. This, naturally, means that multiple outputs, \textit{i.e.}, temperature perturbations on bottom-wall segments, need to be applied to the $N$ different control segments at the bottom wall. GB formulates this control problem by simply asking a single DRL agent to generate $N$ outputs. As discussed in Ref.~\onlinecite{belus2019exploiting}, this approach results in the curse of dimensionality on the control space, which makes learning challenging. By contrast, we apply the invariant MARL approach suggested by Ref.~\onlinecite{belus2019exploiting}, which allows to control an arbitrary number of heating segments without resulting in the curse of dimensionality. This has a considerable impact on our ability to rapidly learn an effective control of the RBC system.

    \item GB uses a lattice-Boltzmann method (LBM), parallelized to run on a specific graphics-processing unit (GPU), to help accelerate the simulations. In such a way, GB partially alleviated the curse of dimensionality introduced by controlling a multiple-output DRL system without using a MARL approach. Indeed, the use of a GPU-parallelized LBM allows GB to run a very large number of DRL trial-and-error steps and, therefore, to partially compensate for the very slow and challenging learning arising due to the curse of dimensionality. Note, however, that this is only applicable for narrow channels with a limited number of control values, where the curse of dimensionality remains under control. Moreover, LBM is effective, especially as they are well adapted to the parallel nature of GPUs, but understanding their accuracy can be challenging. Finally, GB's implementation is not publicly released, and may be hardware dependent. By contrast, we use a spectral Galerkin solver running on standard central-processing Units (CPUs). The solver is both open source, massively parallelizable across many CPU cores, has high order and well-understood accuracy (see Sec.~\ref{sec:methodology} for more information). Moreover, we release all our code as open-source materials (see Appendix~A). Our code can, therefore, be used to further evaluate a wide range of domain sizes and non-dimensional parameters in future studies.
\end{itemize}

The structure of our study is as follows. In Section \ref{sec:methodology}, we describe the formulation of the RBC problem including the geometry, equations and boundary conditions adopted (Section II.A.). We then describe the numerical method used to simulate the CFD (Section II.B.), followed by the details of the DRL-based control (Section II.C.). In Section \ref{sec:results}, we proceed to present the results. We compare the RBC obtained without control (baseline) to the case with control following a non-MARL method similar to GB, and also to the case with control following a MARL approach. We discuss learning speed and robustness. We then show that the invariant MARL controller is able to effectively coalesce RBC cells in the case studied. Finally, we discuss this result and its implication for future works and industrial applications in the conclusions in Section \ref{sec:conclusions}.

\section{Methodology} \label{sec:methodology}

\subsection{Formulation of the RBC problem} \label{subsec:formulation_rbc}

\begin{figure*}
\includegraphics[width=0.9\textwidth]{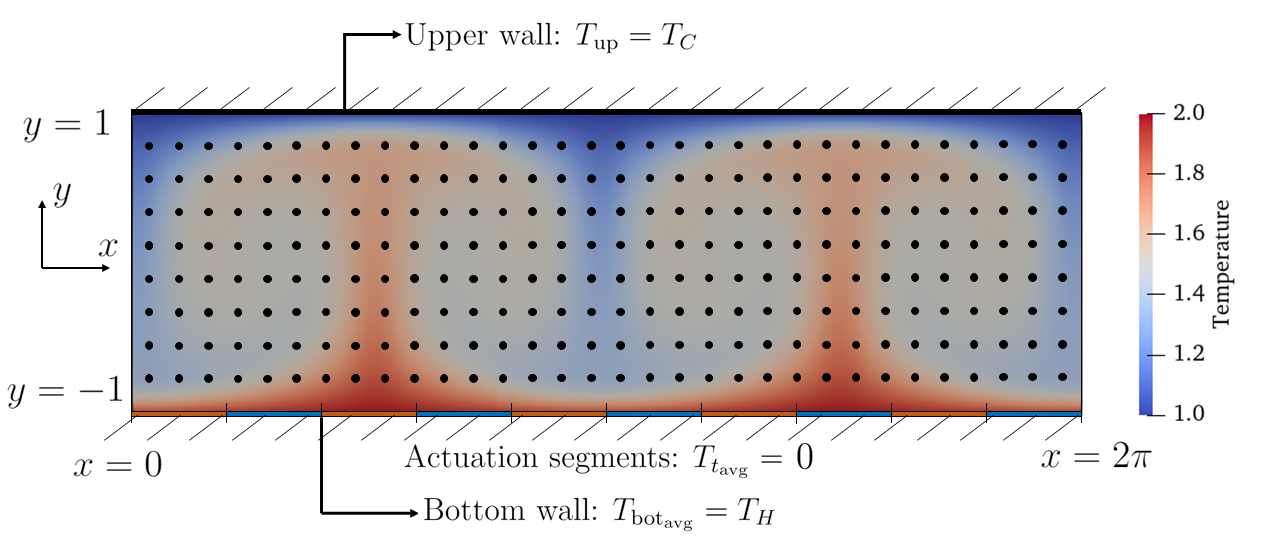}
\caption{\label{setup_sketch} Schematic representation of the domain. Dimensions are normalized by $H$. Temperature is controlled at the bottom wall, which is at an average temperature $T_H$. The upper wall is at a uniform constant temperature $T_C$, so that $T_H>T_C$, generating natural convection and the emergence of the convective cells. Black dots show the position of the observation probes, distributed as a uniform probe-mesh over the domain. Orange and blue segments in the lower wall correspond to the control segments.}
\end{figure*}

The governing equations for this problem are the well-known Navier--Stokes equations. In order to formulate these equations in a form adapted to our problem, we use the Boussinesq approximation, which considers the influence of the density gradients only in the gravitational forces term and assumes that density variations are small compared to velocity gradients. Thus, this formulation neglects density changes in the continuity equation, which then translates into a zero-divergence equation. Therefore, under the Boussinesq approximation, for a non-Newtonian incompressible flow, we obtain the continuity (Eq.~\ref{continuity}), momentum (Eq.~\ref{momentum}), and energy (Eq.~\ref{temperature}) equations in dimensionless form\cite{pan18}:

\begin{subequations}
\label{governing_equations}
\begin{equation}
\nabla \cdot \textbf{u} = 0,\label{continuity}
\end{equation}
\begin{equation}
\frac{\partial \textbf{u}}{\partial t} + \left(\textbf{u}\cdot\nabla\right)\textbf{u} = -\nabla p + \sqrt{\frac{Pr}{Ra}}\nabla^2 \textbf{u} + T\textbf{j}, \label{momentum}
\end{equation}
\begin{equation}
\frac{\partial T}{\partial t} + \textbf{u}\cdot\nabla T = \frac{1}{\sqrt{Ra Pr}}\nabla^2 T, \label{temperature}
\end{equation}
\end{subequations}

\noindent where $\textbf{i}$ and $\textbf{j}$ are the Cartesian unit vectors pointing along the coordinate axes $x$ and $y$, respectively, and the $x$-axis is parallel to the walls. The corresponding velocities for each spatial coordinate are $u(x,y,t)$ and $v(x,y,t)$, respectively, $\textbf{u}(x,y,t) = u\textbf{i}+v\textbf{j}$ is the velocity vector, $t$ is the time, $p(x,y,t)$ is the pressure and $T(x,y,t)$ is the temperature. Along with $Ra$, the Prandtl number $Pr$ is another dimensionless number that governs the dynamics of the flow. These numbers are expressed as follows:

\begin{subequations}
\label{dimensionless_Ra}
\begin{equation}
Ra = \frac{\alpha \left(T_H-T_C\right) g H^3}{\kappa\nu},
\end{equation}
\begin{equation}
\label{dimensionless_Pr}
Pr = \frac{\nu}{\kappa},
\end{equation}
\end{subequations}

\noindent where $\alpha$ is the thermal expansion coefficient of the fluid, $H$ is the normalised domain half height, $g$ is the gravitational acceleration with a downwards vertical direction, \textit{i.e.}, in our case, perpendicular to the walls, $\nu$ is the kinematic viscosity of the fluid, and $\kappa$ is the thermal diffusivity. In our setup, the flow is confined between the bottom hot wall, at a mean temperature $T_H = 2$, and the cooler upper wall, at a uniform constant temperature $T_C = 1$. The former temperature ($T_H$) is also uniform in what we call the `baseline' case (i.e., when no control is applied), but it varies with spatial position in what we call the `controlled' case, as discussed below. Both walls are separated by a distance $2H$ and the no-slip boundary condition is applied at both walls. Periodic boundary conditions are applied at the lateral ends of the domain which has a normalised width $L=2\pi H$. Figure \ref{setup_sketch} provides a schematic representation of the RBC domain.

During the entire study, the Prandtl number employed is $Pr = 0.71$, which corresponds to the Prandtl number of air. As described in the introduction (Sec.~\ref{introduction}), $Ra$ is the ratio of the time scales associated with the thermal transport due to diffusion and convection. This implies that flows with a high $Ra$ are more prone to instabilities due to buoyancy-driven convection. The $Ra$ used in this work is $Ra = 10^4$, which is above the critical $Ra_c$ when no control is applied in the flow.

Another very important non-dimensional number in heat-transfer problems, which is closely related to $Ra$, is the Nusselt number ($Nu$), defined as the ratio of the fluxes due to convective heat transfer to conductive heat transfer. For comparison purposes in the results section of the article we will adopt the same definition as in Ref.~\onlinecite{beintema2020controlling}:

\begin{equation}\label{eq:Nu}
Nu = \frac{\overline{q}}{\kappa\left(T_H-T_C\right)/H},
\end{equation}

\noindent where $\overline{q(t)}$ is the time-averaged heat flux in the domain given by

\begin{equation}\label{eq:q}
\overline{q}=\langle \overline{vT} \rangle_x - \frac{1}{\sqrt{RaPr}} \frac{\partial\langle \overline{T} \rangle_x}{\partial y}.
\end{equation}

\noindent Here the brackets, $\langle\bullet \rangle_x$, indicate averaging with respect to the indicated spatial direction and the overbar, $\overline{\bullet}$, indicates average in time. A description of the method of derivation of Eq. \ref{eq:q} is given in Appendix B.

\subsection{Numerical method for the CFD} 
\label{subsec:num_method_CFD}

The implementation of the governing equations \eqref{governing_equations} is based on a spectral Galerkin version of the method developed by Kim, Moin and Moser \cite{kim_moin_moser_1987} for direct numerical simulations of turbulent flows. In this method the pressure is eliminated and the 2D Navier--Stokes equations are basically reduced to the continuity equation and a fourth-order equation for the wall-normal velocity component. Adding the energy equation, the three scalar equations that are solved are Eqs. \eqref{continuity}, \eqref{temperature}, as well as:

\begin{equation}
    \frac{\partial \nabla^2 {v}}{\partial t} = \frac{\partial^2 H_x}{\partial x \partial y} - \frac{\partial^2 H_y}{\partial x^2}  + \sqrt{\frac{Pr}{Ra}} \nabla^4 {v}  + \frac{\partial^2 T}{\partial x^2}, \label{eq:momentuny}
\end{equation}
where $\bs{H} = (\bs{u} \cdot \nabla) \bs{u}$ is the convection vector. Equation \eqref{eq:momentuny} is implemented with the four boundary conditions $v(x, \pm 1) = v'(x, \pm 1) = 0$, where the first two are due to no slip, whereas the two latter follow from the continuity equation. Periodic boundary conditions on the $x$ direction are applied between the left and right domain boundaries, and this is implemented through the choice of the numerical elements and basis functions, as described below.

The three scalar equations \eqref{eq:momentuny}, \eqref{continuity} and \eqref{temperature} are implemented using a highly-accurate spectral Galerkin \cite{shenbook} discretization in space and a third-order implicit/explicit (IMEX) Runge-Kutta method \cite{ascher97} for the temporal integration. The Galerkin method makes use of tensor product basis functions constructed from Chebyshev polynomials for the wall-normal direction and Fourier exponentials for the periodic direction. The boundary conditions in both directions are built into the basis functions and as such enforced exactly. For the wall-normal direction this requires the use of composite Chebyshev polynomials\cite{shenbook}, whereas the Fourier exponentials for the $x$-direction are already periodic. Since the continuity equation cannot be used to find $u$ for Fourier wavenumber 0, we solve for this wavenumber the momentum equation in the $x$ direction. All other unknowns are closed through Eqs. \eqref{eq:momentuny}, \eqref{continuity} and \eqref{temperature}. The convection terms $\bs{H}$ and $\bs{u} \cdot \nabla T$ are computed in physical space after expanding the number of collocation points by a factor of $3/2$ in order to avoid aliasing. For $\bs{H}$ we use the rotational form $\bs{H} = -\bs{u} \times (\nabla \times \bs{u})$, with the remaining $1/2 \nabla \bs{u} \cdot \bs{u}$ absorbed by the pressure, and for temperature we use the divergence form $\bs{u} \cdot \nabla T = \nabla \cdot \bs{u}T$.

The code is implemented using the open-source spectral Galerkin framework `shenfun'~\cite{mortensen_joss}, where equations can be automatically discretized through a high-level scripting language closely resembling the Mathematics. The Navier--Stokes solver has been verified by reproducing the growth of the most unstable eigenmode of the Orr-Sommerfeld equations over long time integrations. The Navier--Stokes and Rayleigh--Bénard solvers are distributed as part of the shenfun software, and there is a demonstration guide published in the documentation \cite{shenfun}, which also provides a much more detailed description of the numerical method. Links are provided in Appendix A, and we refer the reader curious of all the technicalities of the CFD solver implementation to the resources detailed therein.

The observation probes are distributed over the domain as a uniform probe-mesh as can be seen in Fig. \ref{setup_sketch}. There are $32 \times 8$ probe-mesh points in the $x$ and $y$ directions, respectively. The number of quadrature points in the solution approximation used by the spectral Galerkin solver is $64 \times 96$, with Fourier and Chebyshev basis functions in the $x$ and $y$ directions, respectively. The resulting solution is thus spatially continuous, and the observations are then evaluated from these global, continuous spectral Galerkin functions at the uniform-mesh probe locations. These observations are used in the DRL-based control methodology which we describe next.

\subsection{Control Methodology with DRL}

Using the numerical methods described in the previous section, the flow is solved starting from an initial condition of a constant adverse temperature gradient in the vertical direction. The simulation is performed from this state, leading to the RB instability to occur, which develops into a two-cell configuration that becomes ultimately stationary. No control is applied up to this point, and this constitutes a `baseline' starting state upon which control can be applied. The baseline is saved and then loaded at the start of each episode in the learning phase. The DRL control is performed using the `Tensorforce' \cite{tensorforce} framework, from which we use the PPO algorithm.

As mentioned in the Sec.~\ref{introduction}, the interaction between the agent and the RBC environment occurs through three fundamental channels of communication: the state or observation $s_t$, reward $r_t$ and action $a_t$, where the subscript $t$ is the time step of the CFD. The state $s_t$ consists of sets of time-averaged velocity and temperature $(\tilde{u}_t,\ \tilde{v}_t,\ \tilde{T}_t)$ measured by computational `probes' distributed across the domain in a uniform mesh as seen in Fig.~\ref{setup_sketch}. The averaging is performed over the previous $4$ time steps (where the tilde over the values above represent this average). The observations are flattened into one single-dimensional array and then fed to the agent's neural network input layer. The way in which the observations are passed to the agent differs in the single-agent and multi-agent frameworks, and will be described in the following.

The second channel of communication from the environment to the agent is the reward, $r_t$, which is the desired parameter to be optimized. The goal of the control is to reduce the convective effects, \textit{i.e.}, reduce the intensity of the RBC cells. Since $Nu$ is an indicator of the magnitude of the convective heat transfer, the reward function is set to minimize $Nu$. Specifically, we aim to minimize the instantaneous Nusselt number $Nu_{\rm{inst}}$, which is a function of the instantaneous heat flux $q(t)$. The expression for $q(t)$ is derived in Appendix B. This results in:
\begin{equation}
Nu_{\rm{inst}} = \frac{q(t)}{\kappa\left(T_H-T_C\right)/H}.
\label{Nu_inst}
\end{equation}

While our goal is to reduce $Nu$, DRL algorithms are formulated as an optimization problem that tend to maximize a reward value. Therefore, reward is taken as the negative value of the $Nu_{\rm{inst}}$. This is then followed by a translation and scaling to obtain reward values between $0$ and $1$ (since Tensorforce's PPO algorithm works better within this range), \textit{i.e.}:
\begin{equation}
r_t = m\left(n - Nu_{\rm{inst}}\right),
\label{reward_ini}
\end{equation}

\noindent where $n=2.67$ and $m=1$ are the shifting and scaling, respectively, to set the reward in the range of $[0, 1]$ during the training.

The applied control actuation is a perturbation $T'_t$ of the temperature boundary conditions on the lower hot wall. To apply this in practice, we adopt a distributed actuation framework where the lower wall is divided into $N$ segments and $N$ actuation values are applied independently, one for each of the segments. The raw $a(j)_t,\ j = 1 \ldots N$ that are generated by the Tensorforce agent are non-dimensional scalar values in the interval $[-1, 1]$. These are then transformed into the temperature values applied on each segment $T(j)_t,\ j = 1 \ldots N$ using a shifting and normalisation process that ensures that the average temperature at the bottom wall is kept constant: first, we seek to maintain a constant mean value of temperature, $T_H$, at the lower wall. Changing $T_H$ would modify the regime of the RBC flow characterised by the $Ra$ (Eq. \ref{dimensionless_Ra}), and would thus change the driving mechanism of the instability. Keeping a constant mean $T_H$ prevents this. Next, since the $N$ values of the control temperature are generated independently of each other, a shifting of the control temperature is performed to preserve the lower-wall mean temperature to be equal to $T_H$. The shifted values of temperature, which we denote as $T'_t(j)$, are given as: 
\begin{equation}
\label{control_equations_1}
T'_t(j) = a_t(j) - \frac{\sum_{i=1}^{N}a_t(i)}{N}
\end{equation}
Next, in order to prevent non-physical temperature perturbations, we enforce a limit on the final values of the actuations $T(j)_t < |0.75| \ \forall \ j$ by performing a normalisation:
\begin{equation}
\label{control_equations_2}
T_t(j) = \frac{T'_t(j)}{\max_{j}\left(1,|T'_t|\right)/C}
\end{equation}
\noindent where $C$ is a scaling constant. Therefore, Eqn. (\ref{control_equations_1}) implements the mean subtraction to make sure the temperature change applied by the controller does not change the average temperature at the bottom wall, while Eqn. (\ref{control_equations_2}) makes sure that the maximum absolute value of the temperature change applied is within $0.75$, even in the case where the mean subtraction may lead to one of the $T'_t$ values having a magnitude slightly larger than $1$.

\begin{figure}
\centering
\begin{subfigure}[b]{0.45\textwidth}
\centering
\includegraphics[width=\textwidth]{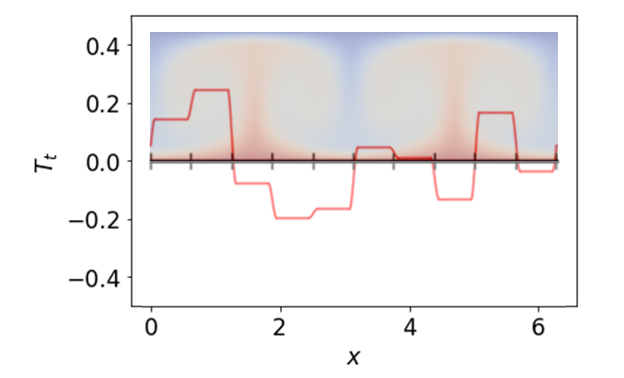}
\label{actions_sketch_global}
\end{subfigure}
\hfill
\begin{subfigure}[b]{0.45\textwidth}
\centering
\includegraphics[width=\textwidth]{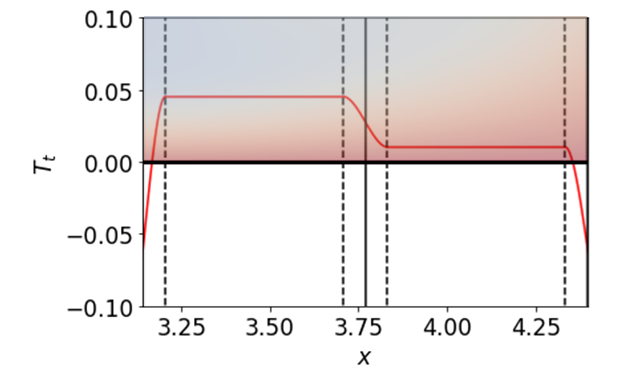}
\label{actions_sketch_zoom}
\end{subfigure}
\hfill
\caption{Schematic representation of how the control $T_t$ is applied to the lower boundary. a) Global representation and b) zoom in the transition region between control segments. Observe that $\sum_{i=1}^{N}T_t(i)=0$, \textit{i.e.} the mean temperature at the bottom wall is not modified by applying the temperature deviations described by $T_t$ to the initially uniform bottom wall temperature $T_H$. Note that the temperature profile plotted in the red line is for illustration purposes, and does not represent a simulation result.}
\label{fig:three graphs}
\end{figure}

Finally, to avoid nonphysical sharp discontinuities in the spatial variation of actuations across the segments, the left and right $10 \%$ of each segment are smoothed relatively to their immediate neighbors following a cubic transition function. In such a way, the cubic function transitions from $T_t(j)$ to $T_t(j+1)$ in a continuous and differentiable way. Fig.~\ref{fig:three graphs} shows a schematic representation of how the temperature control is applied as a whole. An illustration of the cubic transition is shown in the zoomed Fig.~\ref{fig:three graphs}(b).

\begin{figure}
\begin{center}
    \includegraphics[width=0.8\textwidth]{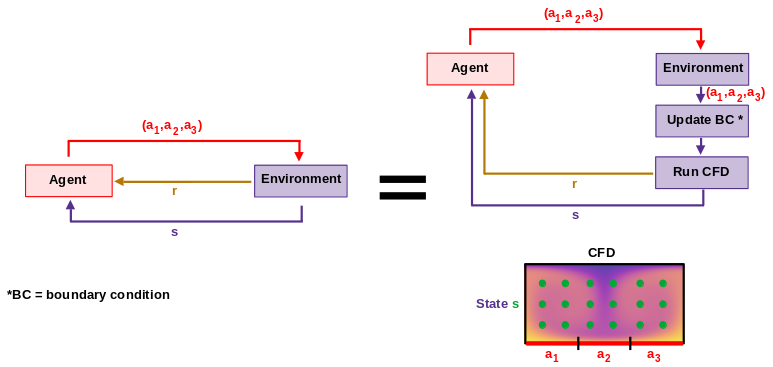}
\end{center}
\caption{Illustration of the single-agent-reinforcement-learning (SARL) control setup. In the SARL setup, the agent controls the environment as a whole in a single step. This means that the dimensionality of the action is equal to the number of control segments, which results in a high-dimensional control space and leads to the curse of dimensionality, as discussed in Ref.~\onlinecite{belus2019exploiting}. Note that, in the present illustration, we draw a thinner channel with only 1 RBC cell and 3 control segments; this is purely for the purpose of illustrating the SARL method, since the actual channel used here is wider and has more control segments, as described above.}
\label{fig_sarl_setup}
\end{figure}

While the general DRL setup we just described is perfectly valid from a DRL point of view, and it is actually implemented and referred to as the `single-agent-reinforcement-learning' (SARL) setup in the following, as illustrated in Fig. \ref{fig_sarl_setup}, it does not exploit the intrinsic structure of the RBC problem. Indeed, the physical mechanism driving RBC does not depend \textit{per se} on the position along the wall direction (\textit{i.e.} along the $x$ axis). This, combined with the periodic boundary condition we implement in the CFD solver on the left and right ends of the domain, implies that the DRL-and-RBC setup is globally invariant along the $x$ direction. This invariance can, therefore, be exploited through implementing a MARL approach, as detailed in Ref.~\onlinecite{belus2019exploiting}, and previously discussed.

In the following, we proceed to define the `MARL-DRL' setup based on modified versions of the states, rewards and actions described thus far. We first regard the $N$ segments introduced previously as separate `pseudo-environments', each with their own independent state, action, and reward channels of communication with the PPO agent. Since our RBC setting is invariant in the $x$ direction, each trajectory in the phase space obtained for each pseudo-environment provides knowledge about the dynamics to control to the DRL agent. As discussed in Ref.~\onlinecite{belus2019exploiting}, this can drastically speed up and improve the quality of the training and learning.

We make a note here regarding the number of pseudo-environments $N$, which is used is a metaparameter of the MARL setup. Typically, $N$ needs to be chosen so that the following needs are balanced:
\begin{enumerate}
    \item The segments should be small enough to ensure sufficient spatial granularity in the control signal to be able to actually control the flow. As a consequence, there should be at least a few control segments per spatial feature of the flow. In our baseline case, given that there are 2 convection cells of length scale $\approx \pi$, we should have at least a few control segments per $\pi$ distance along the $x$ axis.
    \item Despite the fact that MARL is efficient at learning even in the case of many control segments~\cite{belus2019exploiting}, the number of segments should not be too large to overcome certain practical limitations:
    \begin{itemize}
        \item A large number of segments will make the control output discontinuous as a large number of points, requiring the use of more aggressive smoothing functions.
        \item Each pseudo-environment corresponding to each segment will require its own stream of data to be passed to and received from the agent. From a computational viewpoint, this increases the cost of the DRL deployment.
        \item Neural networks tend to learn best when learning from uncorrelated data. This means that, ideally, feeding similar data many times to the DRL agent should be avoided. Therefore, providing segments large enough so that the view on each of them is significantly different from the views on the others is beneficial for the quality of the learning.
        \item Segments that are too small in size can produce an overall temperature profile (which is the control output), which is highly non-uniform. As a consequence, a Fourier decomposition of the corresponding temperature profile in space will consist of elements with a high wavenumber energy content. These high wavenumber energy content elements can make it more challenging to obtain good stability of the CFD simulation. 
    \end{itemize}
\end{enumerate}

Considering the points above, we choose a value of $N=10$ (Table~\ref{parameters}) as a reasonable trade-off. We note that this is the analogous (in space) to the considerations that are presented in Ref~\onlinecite{rabault2019accelerating} when choosing the action duration and episode length of the DRL algorithm (\textit{i.e.} in time).

In the following, we resume the description of how the states, actions and rewards communication channels are modified in the MARL-DRL setup. The state $s_t$ of each pseudo-environment is defined as the global set of observations whose order has been rearranged in such a way that the observations from the probes directly above its respective control segment are in the center of the observation list. This `recentering' is further elaborated in the description of the schematic of Fig.~\ref{fig_marl_setup} below. 

In a similar way, the reward function, $r_t$, that is provided to each MARL environment, is slightly modified. Separate rewards are computed from each MARL pseudo-environment, and now two values of $Nu$ contribute to the computation of each local MARL reward. The first is the global $Nu$ which is equal to $Nu_{\rm{inst}}$  (Eq.~(\ref{reward_ini})). This part of the reward accounts for the main goal of the agent to optimize the flow over the whole channel. This gives incentive to each pseudo-environment to improve the global flow state (rather than behaving in an egoistically-greedy way). The second contribution to the reward is from a local Nusselt number, $Nu_{\rm{loc}}$, which is calculated for each specific pseudo-environment in the column of observation probes immediately above its control segment. This local contribution incorporates information about the local effects of the actions on the flow to the reward. This, in turn, allows to provide more reward granularity to the agent during the training, by generating a local indication of the quality of the control. A constant weighting factor, $\beta=0.0015$, is used to quantify the contribution of $Nu_{\rm{loc}}$ and $Nu_{\rm{inst}}$ to the total reward $r_t$ for each pseudo-environment. The value of $\beta$ appears to be low, but this is only due to the large difference in magnitudes that exists between $Nu_{\rm{loc}}$ and $Nu_{\rm{inst}}$. This large difference in magnitude arises from the difference in the geometric width used to compute $Nu_{\rm{loc}}$ and $Nu_{\rm{inst}}$ (since we use $N=10$ segments, each segment has a width factor of $1/10$ with respect to the total domain width). We choose $\beta$ such that $Nu_{\rm{loc}}$ has approximately a $10\%$ contribution to the total reward. The final form of the reward function for each MARL pseudo-environment reads as follows:
\begin{equation}\label{eq:total_reward}
r_t = m(n- (1 - \beta) Nu_{\rm{inst}} - \beta Nu_{\rm{loc}}),
\end{equation}

\noindent with values of $n$ and $m$ as defined for equation (\ref{reward_ini}). 

A schematic representation of the MARL-based control methodology is presented in Fig.~\ref{fig_marl_setup}. In order to make the figure easier to read, only three representative pseudo-environments are shown there instead of $N=10$. Each pseudo-environment corresponds to a localized region of the physical domain. Note that for ease of explanation, in this plot, we deviate from the notation convention used in the rest of the paper: the subscripts for $s$, $a$ and $r$ are now the environment index $j=1,2,3$. We define $s_j$ as the values of the observation from the probes directly above each control segment, $r_j$ as the total reward from each pseudo-environment (Eq. \ref{eq:total_reward}) and $a_j$ as the control actions applied by each MARL pseudo-environment. The top diagram represents the control methodology. At the start of the cycle, the $a_j$ are merged and communicated to the first pseudo-environment ($j=1$), which is the sole pseudo-environment that runs the whole CFD simulation, while the others pseudo environments wait until CFD simulation is finished. The CFD simulation is run for a time equal to the action duration (Table \ref{parameters}). Once the action is finished, the entire set of observations $s=(s_1,\ s_2,\ s_3)$ is retrieved from the observation probes, and communicated to all the pseudo-environments. Then, in the `Merge results' block, each pseudo-environment executes the `recenter()' function. In `recenter()', a rearranging of the order of $s_j$ is performed in each pseudo-environment, so that the $s_j$ corresponding to pseudo-environment $j$ is moved to the middle of the observation vector. We denote the recentered observation vectors as $s'_j$. The $r_j$ are obtained from the $Nu_{\rm{loc}}$ and $Nu_{\rm{inst}}$ computed using the local $s_j$ and the global $s$ respectively. This is done using Eq. \ref{eq:total_reward}, and it is represented for brevity as a function of $g$ in Fig.~\ref{fig_marl_setup}. The $s'_j$ and $r_j$ are then passed to the respective MARL agents, which output the actions $a_j$ and the cycle is repeated again. The MARL agents are actually implemented as data streams to a repeated DRL agent, which means that all MARL agents share the exact same policy and neural networks. This is how the control invariance across the domain is implemented in the MARL case, \textit{i.e.}, by re-using the exact same agent multiple times across the domain, similar to what is described in Ref.~\onlinecite{belus2019exploiting}.

\begin{figure}
\begin{center}
    \includegraphics[width=0.7\textwidth]{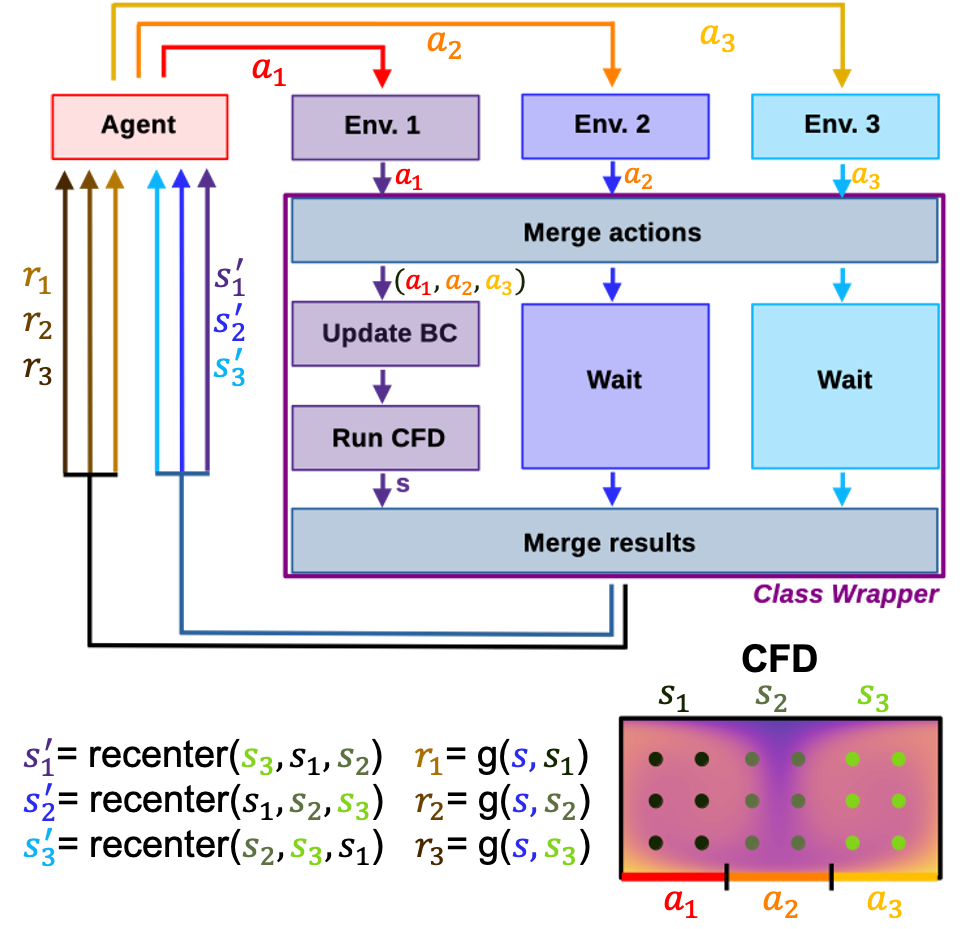}
\end{center}
\caption{Schematic of the MARL control setup, where the same agent controls local regions of the domain as if they were separate environments. This allows to i) re-use the same policy knowledge across the domain, ii) get a local reward information during training. This, in turn, allows to alleviate the curse of dimensionality on the action space dimension when controlling systems with invariants, as described in Ref.~\onlinecite{belus2019exploiting}. A code wrapper is needed to ensure synchronization of the parallel local environments with the CFD simulation. Note that for illustrative purposes, the bottom right diagram shows only three pseudo-environments instead of $N$.}
\label{fig_marl_setup}
\end{figure}

\begin{table}[]
\begin{tabular}{lr}
\textbf{Parameter}      & \textbf{Value}  \\
\hline
Domain size $L\times H$ & $2\pi\times 2$ \\
Galerkin Modes             & $96\times 64$  \\
Time step               & $0.05$ \\
$Pr$                    & 0.7     \\
$Ra$                    & $10^4$        \\
$\beta$ & 0.0015   \\
Action scaling factor, $C$      & 0.75   \\
Number of observation probes    & $8\times32$  \\
Number of CFD episodes              & 350       \\
Number of action steps per episode   & 200 \\
Number of control segments, $N$ & 10 \\ 
Baseline duration          & 400 time units \\
Action duration  & 1.5 time units \\
Episode duration           & 300 time units \\
\hline
\end{tabular}
\caption{Parameters used in the CFD simulation and DRL control. Note that for MARL cases, we make a distinction between `MARL episodes' (of which we have $N=10$ per CFD simulation that lasts for one episode duration) and `CFD episodes' (of which we have one per CFD simulation that lasts for one episode duration). In general, episodes are defined by the duration of performing a set number of control actuations on the environment, before terminating the simulation of the environment. In a CFD episode, one actuation is a single set of $N$ control actions for the $N$ pseudo-environments. In a MARL episode, one actuation is one control action performed on one pseudo-environment. Given in the table that the number of actuations per episode is 200, this means there are 200 sets of actuations for a single CFD episode, which corresponds to a total of $200\times N$ actuations performed from the MARL actors as a whole.}
\label{parameters}
\end{table}

Having described the setup of SARL and MARL, we now proceed to the training process. The main parameters used for the DRL setup are summarized in Table \ref{parameters}, and the hyperparamters used for the definition and training of the agent are in Table \ref{table:agent_parameters}. A total of 350 CFD  episodes are performed for the training run in SARL, with 200 actuation updates per episode. Each actuation consists of a set of $N$ actions or temperature control values for the $N$ segments. In SARL, the policy update is performed every `batch size’ number of episodes, which in our case is set to be 20. In MARL, we make a distinction between `CFD episodes' and `MARL episodes'. The `CFD episode' is when the CFD simulation is run for 200 actuations, as in the SARL case. Each actuation comprises $N$ actions applied locally by each MARL pseudo-environment. Thus, each MARL pseudo-environment, in the course of one CFD episode, applies 200 control actions, which we define as a single MARL episode. Given that there are $N$ MARL pseudo-environments, there are therefore 350$\times N$ MARL episodes in one training run for MARL. The batch size for MARL is 20 MARL episodes, or 20$/N$ CFD episodes. In the discussion of the results section, in the context of MARL, the word `episode' refers to `CFD episode'.

Training in both SARL and MARL is performed with a discount factor of $\gamma = 0.99$. Wall-clock training time for 350 episodes in SARL and 350 CFD episodes in MARL is around 34 hours using a single core on a modern workstation (the wall clock duration is approximately the same in both cases, as CFD is the dominating computational cost). The duration of each episode in CFD time units is action duration $\times$ number of actions per episode $=300$ time units (Table \ref{parameters}). This time is enough to capture the dynamics of the flow in the transition from the baseline state to the optimized controlled flow. Within the duration of one action (i.e., within a single control step), the control temperature applied to each control segment is established by the DRL agent through equations (\ref{control_equations_1}) and (\ref{control_equations_2}). This control step is chosen to be short enough so that the agent can actuate under fast changes of the state of the flow, but also long enough so that the flow can develop according to the actuation performed, similar to what is recommended in \citet{rabault2019artificial}.

\begin{table}[]
\begin{tabular}{lr}
\textbf{Hyperparameter}                                            & \textbf{Value} \\
\hline
Batch Size for SARL                                                   & 20 episodes           \\
Batch Size for MARL                                                   & 20 MARL episodes           \\
Learning Rate                                                 & $10^{-3}$      \\
Entropy Regularisation Factor                                 & 0.01           \\
Number of hidden layers (baseline network and policy network) & 2              \\
Number of units per hidden layer                              & 512            \\
Activation function                                           & tanh           \\
Optimizer                                                     & Adam         \\
\hline
\end{tabular}
\caption{Hyperparameters used in the definition and training of the DRL agent. The hyperparameters are common to SARL and MARL.}
\label{table:agent_parameters}
\end{table}

\section{Results and discussion}\label{sec:results}

\begin{figure}[t]
\includegraphics[width=155mm]{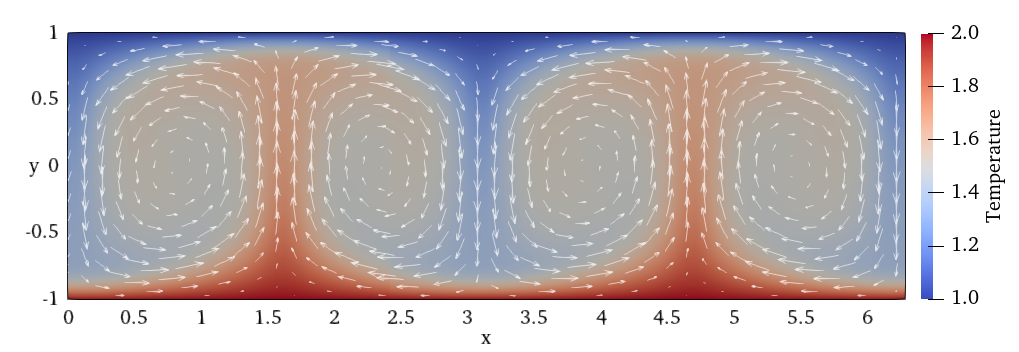}
\caption{\label{fig:3_baseline_fields} Instantaneous temperature (background color) and velocity (arrows) fields from the end of the baseline simulation. This is the baseline steady state used at the start of each new episode, displaying two convective cells ($Nu = 2.68$) (see also the corresponding multimedia data from Appendix C).}
\end{figure}

In the following section we analyse and discuss the results obtained. In the first place, at the start of the baseline convergence simulation, three convective cells appear. However, this is an unstable state that lasts for just a few time units and the final baseline, which includes a two-cell configuration, is achieved as visible in Fig.~\ref{fig:3_baseline_fields} (see also the corresponding multimedia data from Appendix C). A full video of the baseline simulation can be seen in the link provided in Appendix C.

\begin{figure}
\includegraphics[width=110mm]{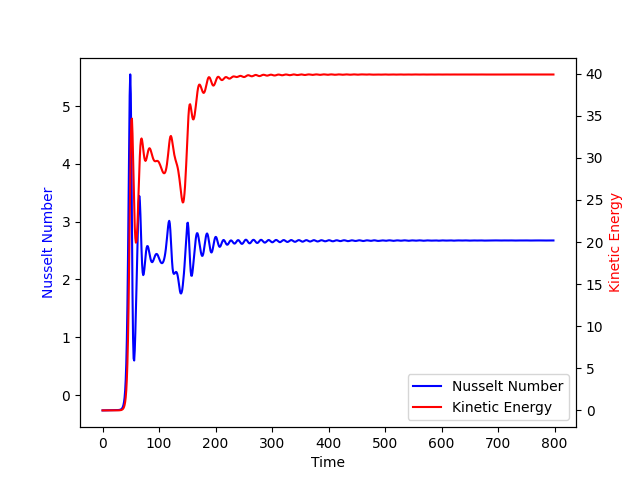}
\caption{\label{fig:2_baseline} Evolution of the global Nusselt number and kinetic energy during the baseline simulation with $Ra=10^4$.}
\end{figure}

Fig.~\ref{fig:2_baseline} shows the evolution of the Nusselt number and the kinetic energy along the duration of the baseline convergence simulation. The flow reaches a stable state after approximately $400$ time units, where the Nusselt number converges to a value of $Nu = 2.68$.

In the baseline flow two convective cells are present, with a total of two counter-rotating vortex pairs. This is different from the baseline layout obtained in GB, where a single rotating vortex was obtained. The differences in flow topology with respect to the results by GB come from the adiabatic lateral walls and the aspect ratio of $1$ used in GB, compared with the periodicity in $x$ and an aspect ratio of $\pi$ used in the present work. Therefore, we can expect different control strategies compared to those obtained in GB.


\begin{figure}[t]
\includegraphics[width=0.7\textwidth]{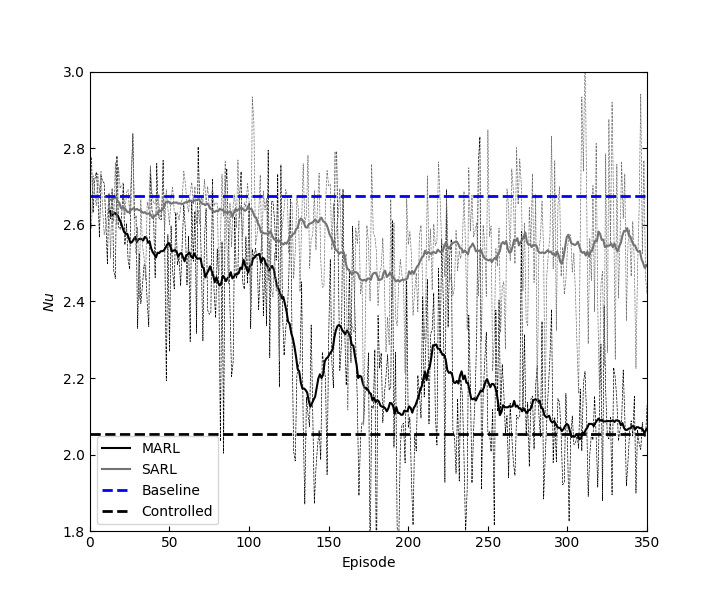} 
\caption{\label{fig:4_vanillaDRLvsMARL} Learning curves obtained in both the SARL and MARL configurations. Note that each episode in the horizontal axis corresponds to a single CFD run, which consists of $N=10$ individual MARL episodes when the DRL algorithm is applied in the MARL configuration. The dotted lines in greyscale are the actual mean rewards obtained per episode, and the full lines are moving averages over 25 episodes that illustrate the learning trends. The horizontal bold dashed lines are drawn for reference, and they represent $Nu$ at the end of the baseline simulation (in blue) and the average $Nu$ from an episode where MARL control is executed (in black). A movie showing control with the trained SARL agent is shown in the link provided in Appendix C.}

\end{figure}

We now present the results obtained from DRL control of the Rayleigh--B\'enard system. To compare the learning in the SARL and MARL frameworks, we plot both their learning curves in Fig. \ref{fig:4_vanillaDRLvsMARL} (see also the corresponding multimedia data from Appendix C). The learning was performed for 350 CFD episodes of 200 actuations each as previously highlighted, starting from the final state of the baseline (Fig. \ref{fig:3_baseline_fields} (see also the corresponding multimedia data from Appendix C)). This means that the same CFD duration is used to train both the SARL and MARL agents. Since CFD simulation is by far the most time- and resource-intensive part of the training process (similar to what was reported in Ref.~\onlinecite{rabault2019artificial}), this means that the wall-clock time and resources used for performing the SARL and MARL trainings are about equal. The time-averaged $Nu$ at the end of the baseline is plotted for reference, corresponding to a value of $2.68$ in Fig.~\ref{fig:4_vanillaDRLvsMARL} (see also the corresponding multimedia data from Appendix C). We also show the time-averaged $Nu$ at the final episode of the MARL, which is equal to $2.07$, corresponding to the dashed horizontal black line.

Fig.~\ref{fig:4_vanillaDRLvsMARL} (see also the corresponding multimedia data from Appendix C) clearly demonstrates that the MARL approach learns much faster and effectively than the SARL approach. This figure shows that, while the SARL approach plateaus after around 150 episodes and displays only very limited $Nu$ reduction, the MARL method learns initially much faster than SARL, and keeps learning for a much longer time until it reaches a far better $Nu$ reduction. To be more specific, at the end of 350 episodes, the MARL approach is able to achieve a reduction in $Nu$ of $22.7\%$, whereas the SARL approach achieves approximately $5.1\%$ reduction. Moreover, a closer look at the last state of the last episodes in the SARL and MARL cases, respectively, shows that while the MARL agent has managed to change the topology of the flow and to reach a coalesced RBC convection cell state (more details below), the SARL approach has only a limited effect on the flow topology (visible from the videos in Appendix C). This implies that, while the MARL agent has learned an effective policy and control law, the agent in the SARL framework has failed to do so. This is very similar to the findings reported in Ref.~\onlinecite{belus2019exploiting}, as previously discussed. The reader curious of more details about the flow dynamics in each case is referred to the videos in Appendix C to see the difference in the controlled-flow configurations. Further insight into the MARL control mechanism is provided in the next paragraphs.

The price to pay for the DRL ability to discover and optimize non-linear control laws is the lack of direct interpretability of the DRL agent policy. Given a trained DRL agent, and in particular its policy network weights, it is difficult to derive a detailed, quantitative understanding of the control law using a bottom up approach, and the DRL policy \textit{per se} is effectively a black box. This contrasts with e.g. linear flow-control methods, where properties such as the eigenmodes and eigenvectors of the associated operators can be used to extract some form of explainability \cite{reddy1993pseudospectra, schmid1994optimal}. However, a higher-level, phenomenological understanding of the DRL agent policy is still obtainable, by applying expert knowledge to the analysis of the control laws exhibited by the DRL agent. Such a phenomenological understanding is what we discuss in the following paragraphs.

\begin{figure}[t!]
\includegraphics[width=76mm]{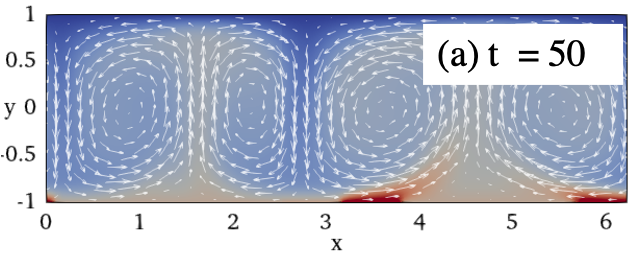}\label{cartoon_a}
\hspace{-1mm}
\vspace{2mm}
\includegraphics[width=76mm]{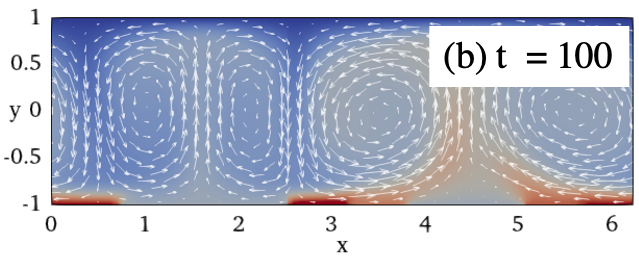}\label{cartoon_b}
\vspace{2mm}
\includegraphics[width=76mm]{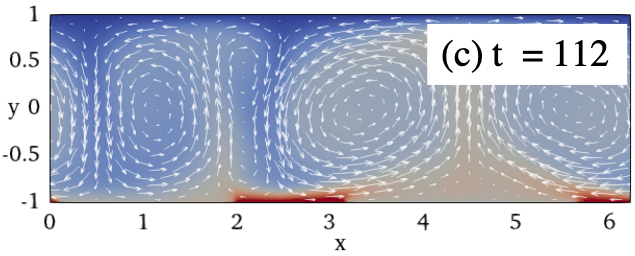}\label{cartoon_c}
\hspace{-1mm}
\includegraphics[width=76mm]{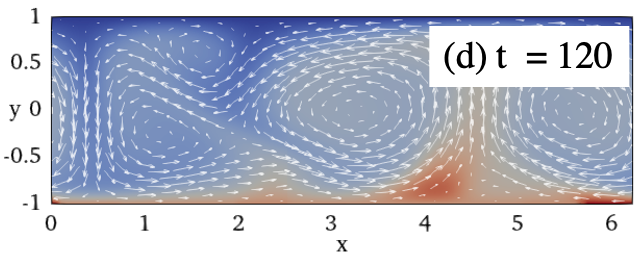}\label{cartoon_d}
\vspace{2mm}
\includegraphics[width=76mm]{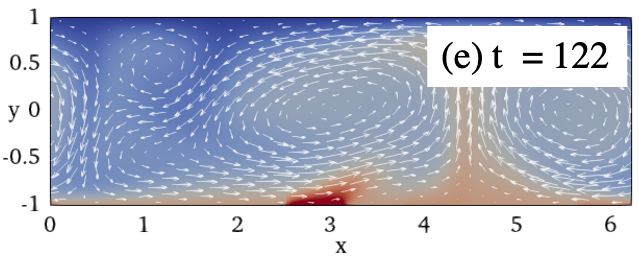}\label{cartoon_e}
\hspace{-1mm}
\includegraphics[width=76mm]{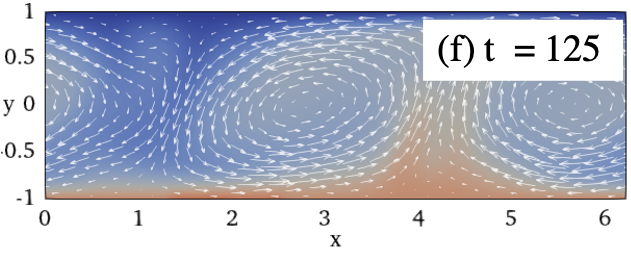}\label{cartoon_f}
\vspace{2mm}
\includegraphics[width=76mm]{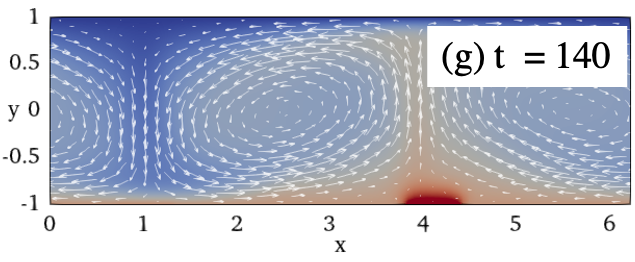}\label{cartoon_g}
\hspace{-1mm}
\includegraphics[width=76mm]{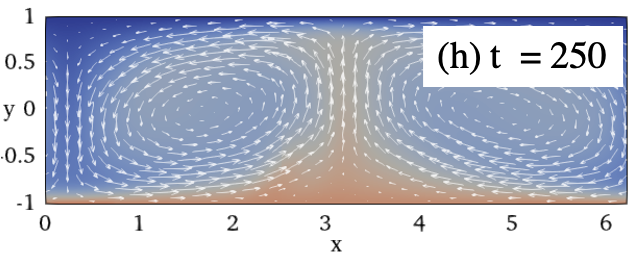}\label{cartoon_h}
\vspace{2mm}
\includegraphics[width=80mm]{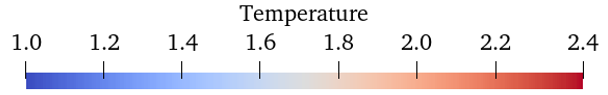}
\caption{Sequence of temperature (background color) and velocity (arrows) fields showing the dynamics of the control action taken by the trained agent during a single episode without exploration noise (deterministic, or evaluation, mode). Starting from the baseline double-RBC-cell flow, the agent destabilizes the flow, leading to coalescence into a single cell. This cell is then controlled until it reaches a stable configuration. Each frame in the current figure corresponds to the snapshots of the flow field at time instants marked with red dot symbols in Fig. \ref{fig:6_Nu_main_episode}, over the different phases of flow development. (see also the corresponding multimedia data from Appendix C)}
\label{fig:6_mechanism}
\end{figure}

More specifically, different phases are clearly visible when the DRL agent controls RBC, as highlighted in Fig. \ref{fig:6_mechanism} (see also the corresponding multimedia data from Appendix C). Starting from the baseline flow, the DRL agent starts by weakening the RBC cells through modulation of the bottom wall temperature. We will refer to this destabilization of the baseline regime as the Phase I. The destabilization is performed by the agent through applying i) increased heating on the regions of the bottom plate where a falling plume of low temperature fluid is observed, and ii) decreased heating (or relative cooling) on the regions of the bottom plate where a rising plume of high temperature fluid is observed. This is visible in Fig. \ref{fig:6_mechanism} (a) and (b) (see also the corresponding multimedia data from Appendix C). This, in practice, weakens the driving mechanism of the baseline RBC cells, by producing excess buoyancy in the downwelling part of the RBC plume. As a result, the control on Phase I reduces the excess of buoyancy in the rising part of the RBC plume, hence, opposing the buoyancy driven flow motion.

\begin{figure}[t]
\includegraphics[width=140mm]{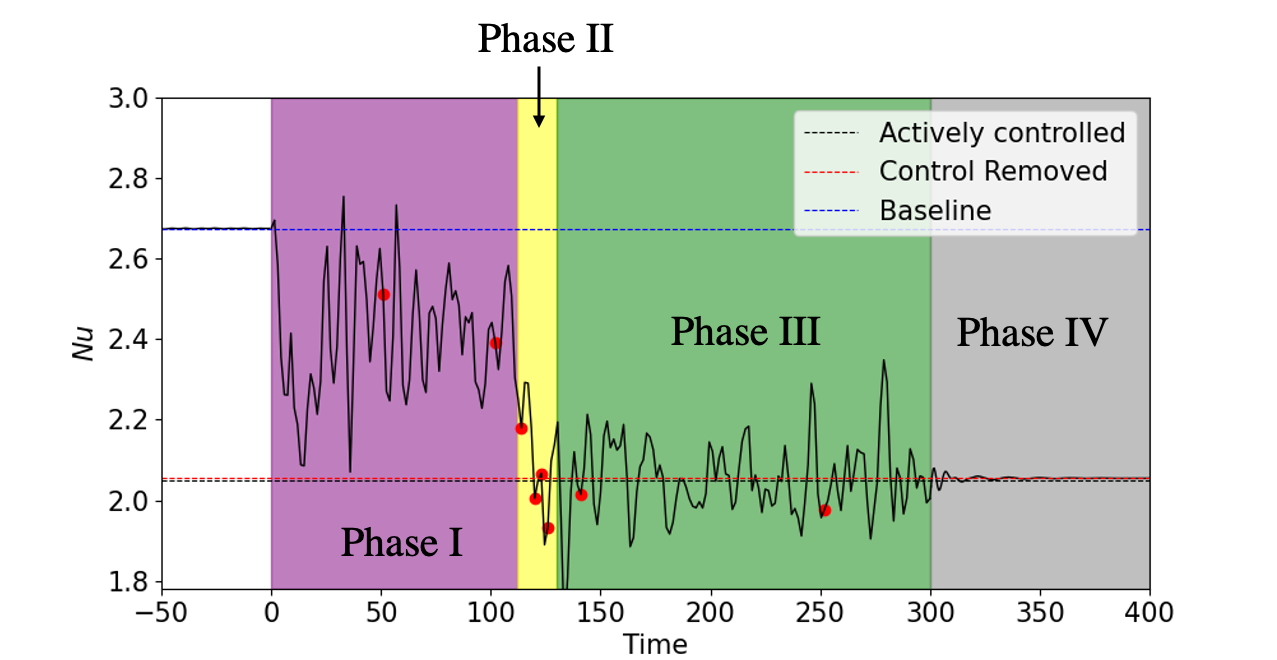} 
\caption{Evolution of the Nusselt number showing the control achieved by the DRL agent in a single deterministic run (\textit{i.e.} without exploration noise in the DRL agent policy). The shaded area shows the different phases of the flow evolution. The red point markers are the time instants at which the frames in Fig. \ref{fig:6_mechanism} are sampled. The black horizontal dashed line is the mean $Nu$ of the actively controlled single-cell configuration from the current episode. The red horizontal dashed line is the $Nu$ achieved at the end state of the system after the control is removed and the naturally stable single-cell RBC configuration is observed (corresponding to the flow visible in Fig. \ref{fig:7_removeControl_field}).}
\label{fig:6_Nu_main_episode}
\end{figure}

This opposition to and weakening of the baseline flow results in an unstable situation, where the double RBC cells are broken. We recognize this transition as the Phase II and it can be observed in Fig.~\ref{fig:6_mechanism} (c)-(f) (see also the corresponding multimedia data from Appendix C) and in Fig. \ref{fig:6_Nu_main_episode}. This Phase II ends with a recombination of the two initial convective cells into a single coalesced RBC cell, thus, leading to the Phase III. Phase III is distinguished by a single coalesced bubble regime, which is visible in Fig. \ref{fig:6_mechanism} (g) and (h) (see also the corresponding multimedia data from Appendix C).

Therefore, it appears that the DRL agent has learnt to ``navigate'' in the phase space of the problem. While the phase space configuration corresponding to the baseline configuration is stable, it leads to a suboptimal reward. In particular, the DRL agent has found a strategy (applied during the transition Phases I and II) that can bring the system away from the baseline configuration and into the neighborhood of a new topological flow configuration, \textit{i.e.}, the coalesced bubble observed in Phase III, which provides higher reward.

\begin{figure}[t]
\includegraphics[width=140mm]{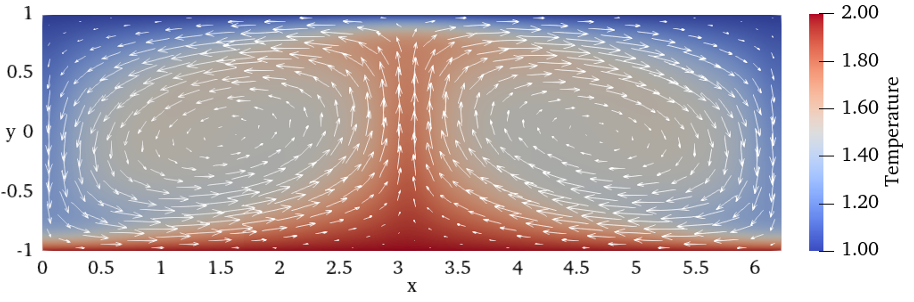} 

\caption{\label{fig:7_removeControl_field} Illustration of the single-cell configuration maintained by the system after the DRL-control is removed and the uniform hot-wall boundary condition re-imposed. Interestingly, once the DRL controller has moved the flow to this new configuration, it is intrinsically stable and yields the best reward, even when left unactuated (see also the corresponding multimedia data from Appendix C).}
\end{figure}

Interestingly, the single coalesced bubble obtained in Phase III is naturally stable once it is established and is in fact providing the optimal reward, even without control. This is what we define as Phase IV in Fig.~\ref{fig:6_Nu_main_episode}, i.e., the flow obtained upon the re-imposition of the lower-wall constant temperature boundary condition (as in the baseline) after the end of episode. A snapshot of this final state is observed in Fig. \ref{fig:7_removeControl_field} (see also the corresponding multimedia data from Appendix C).

\begin{figure}[t]
\includegraphics[width=150mm]{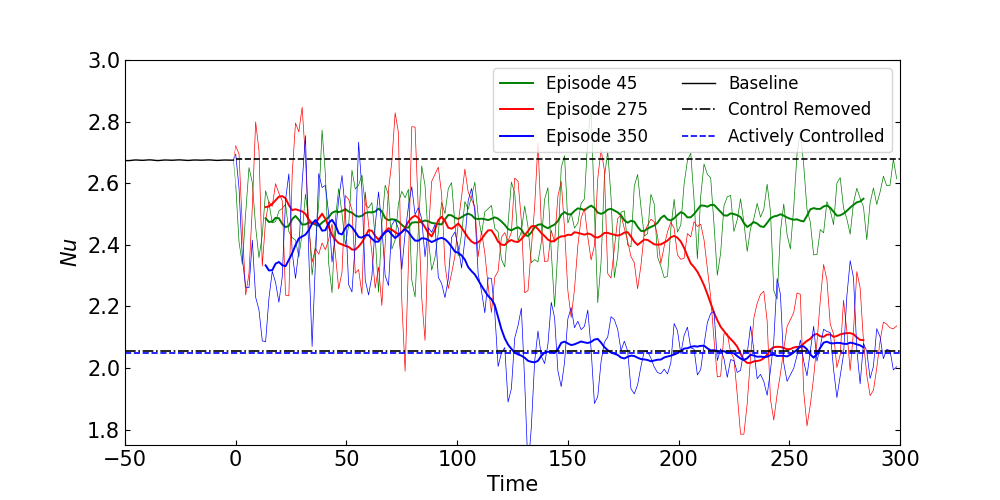}
\caption{\label{fig:5_multipleEpisodes} Evolution of the Nusselt number in three different episodes (episode numbers 45, 275 and 350), chosen from the MARL learning curve in Fig. \ref{fig:4_vanillaDRLvsMARL}. The black horizontal line (dashed for positive times) represents the value of $Nu$ at the baseline. The dash-dotted horizontal line is the $Nu$ number at Phase IV. The control corresponding to episode 45 only leads to a minor change in $Nu$. By contrast, the controls corresponding to episodes 275 and 350 lead to much better $Nu$ reduction, corresponding to the transition to a single coalesced bubble. The difference between episodes 275 and 350 lies in the ability of the controller obtained at episode 350 to reach the coalesced state faster than what is the case at episode 275.}

\end{figure}

Figure \ref{fig:5_multipleEpisodes} shows the Nusselt number obtained through three different episodes at different stages of the learning. While at episode $45$ the DRL agent was not able to coalesce the two convective cells into one, at episodes $275$ and $350$ it has already learned the strategy to perform the coalescence. This is indicated by the large jumps in the Nusselt number at time units $115$ and $215$ for episodes $350$ and $275$, respectively. We can also observe that the agent trained for more episodes (episode $350$) is able to reach the Phase III (single-cell configuration) earlier than the agent trained for fewer episodes (episode $275$). However, once the flow configuration is a single cell, no matter when it is reached, the value of the Nusselt number converges to a minimum of approximately $2.1$. This is related with the fact that the single-cell configuration is a stable state with a higher reward, i.e., lower $Nu$, than the two-cell configuration. If once on Phase III the control is turned off, \textit{i.e.}, in Phase IV, the Nusselt number achieved gradually converges to $2.05$, which is the best $Nu$ we ever observe. Therefore, any small control applied during Phase III is noise (either exploration noise during training, or agent ``uncertainty" during evaluation). The black dash-dotted horizontal line in Figure \ref{fig:5_multipleEpisodes} shows the value of the Nusselt number for Phase IV.

\section{Conclusions}\label{sec:conclusions}

In the present work, we demonstrate effective Rayleigh--B\'enard Convection (RBC) control by using deep reinforcement learning (DRL). Compared with previous works, \textit{e.g.} Ref.~\onlinecite{beintema2020controlling}, we consider a domain with a higher aspect ratio, i.e., a wider domain. This leads to the existence of several rolls or RBC cells, and the need for using more separate actuators at the bottom walls. We show that a naive single agent DRL method fails to learn an effective control policy. This is due to the difficulty presented by having to control several actuators simultaneously, which results in the curse of dimensionality in the control space. By contrast, we demonstrate that leveraging invariant multi-agent reinforcement learning (MARL), which takes advantage of the invariant structure of the underlying RBC control problem, allows to discover an effective control strategy in the case studied. We demonstrate that, in practice, ``multi-agent reinforcement learning is all you need'' to control our multiple-input, multiple-output fluid-dynamics problem. This finding is similar to previous results \cite{belus2019exploiting}, and we expect to keep observing it repeatedly when applying DRL to control physical systems with strong underlying structures and symmetries. Using a MARL approach results in a much faster learning than what was obtained in Ref.~\onlinecite{beintema2020controlling}, and is key to our ability to control wider channels with more actuators.

Going into more details, our MARL controller finds a complex, non-trivial control strategy in the RBC configuration under consideration. Through trial and error, the MARL approach discovers that adequate control of the lower wall temperature profile can destabilize the double RBC cell pattern observed in our baseline flow, and force these RBC cells to coalesce into a single convection cell. This convection cell is, in turn, intrinsically stable, and yields better heat exchange performance, \textit{i.e.}, a lower value of the Nusselt number $Nu$, following the definition of our optimization problem.

The present study is a first application of MARL to RBC control, and we provide a strong framework for further studies through the release of all the codes and scripts as open material. Many interesting questions remain open to discussion in future works, and we expect that more studies of the RBC controllability using MARL will take place. In particular, this framework is well adapted to studying problems such as the effect of the Rayleigh number $Ra$, the channel aspect ratio, the strength of the control authority, etc, on controllability and optimal states of the RBC system. We believe that this work is also a first step towards demonstrating RBC MARL control in both 3D configurations, and in geometries relevant for industrial applications.

\section*{Acknowledgements}

Part of the deep-learning-model training was enabled by resources provided by the National Academic Infrastructure for Supercomputing in Sweden (NAISS). RV acknowledges financial support from ERC grant no. `2021-CoG-101043998, DEEPCONTROL'. Views and opinions expressed are however those of the author(s) only and do not necessarily reflect those of the European Union or the European Research Council. Neither the European Union nor the granting authority can be held responsible for them.

\section*{Data Availability}

The data that support the findings of this study may be generated from the code in the GitHub repository linked to this paper: \url{https://github.com/KTH-FlowAI/DeepReinforcementLearning_RayleighBenard2D_Control.git}.

\section*{Author Declarations}

The authors have no conflicts to disclose.

\section*{Appendix A: Code Release} \label{AppA}

All the codes, scripts, and post-processing tools used in this work are made available on GitHub together with readmes and user instructions, at the following address: \url{https://github.com/KTH-FlowAI/DeepReinforcementLearning_RayleighBenard2D_Control.git}. Reasonable user support will be provided through the issue tracker of the corresponding GitHub repository.

The shenfun CFD case setup is described in great details, including all the numerical implementation considerations, elements chosen, and code walk-through, on the shenfun RBC documentation page: \url{https://shenfun.readthedocs.io/en/latest/rayleighbenard.html}. Note that the conventions used in the present paper and in this documentation page are slightly different: while the present paper uses $x$ as the horizontal (``wall parallel") description and $y$ as the vertical (``wall normal") direction, the documentation page uses the opposite conventions. This has no influence on the CFD results \textit{per se}. The code uses the same conventions as the shenfun documentation, and only the present paper uses separate conventions, for simplicity of the writing and conformity with the literature.

\section*{Appendix B: Heat-flux derivation}\label{AppC}

In this section, we derive the expression for the non-dimensional time-averaged heat flux $q$ as provided in Eq. \ref{eq:q}. We begin with the non-dimensional energy equation (Eq. \ref{temperature}). To this end, we add Eq. \ref{continuity} to obtain:

\begin{equation}\label{eq:appB_1}
    \frac{\partial T}{\partial t} + \nabla \cdot (\textbf{u}T) = \frac{1}{\sqrt{RaPr}}\nabla\cdot\nabla T.
\end{equation}

The above equation is volumetric, and can thus be integrated over the entire volume of the domain. We then apply the Gauss' divergence theorem to obtain:   
\begin{equation}\label{eq:appB_2}
    \int_V \frac{\partial T}{\partial t} \rm{d}V + \int_S (\textbf{u}T) \cdot \mathbf{\hat{n}}\ \rm{d}S = \frac{1}{\sqrt{RaPr}} \int_S \nabla T\cdot\mathbf{\hat{n}}\ \rm{d}S,
\end{equation}

\noindent where $S$ and $V$ are the boundaries and volume of the domain and $\rm{d}S$ and $\rm{d}V$ represent elementary surface and volume units respectively. Also note that $\mathbf{\hat{n}}$ is the unit vector normal to $S$. The second term on the LHS and the term on the RHS represent surface-integrated `flux terms' corresponding to the convective and conductive heat fluxes $q_{\rm{conv}}$ and $q_{\rm{cond}}$ across $S$, respectively.

Since the domain in the current study is two dimensional, integrals in the $z$ direction evaluate to a constant and the surface integrals reduce to line integrals over the domain boundaries. Given that the boundary conditions are periodic in $x$, the component of the flux terms integrated over the vertical boundaries, \textit{i.e.}, from $y=-1$ to $1$, at $x=0$ and $x=2\pi$ sum to $0$. Hence, we are left with the terms integrated over the horizontal boundaries or walls of the domain. The second term on the LHS of Eq.~\ref{eq:appB_2} becomes  
\begin{equation}\label{eq:appB_3}
    \int_{x} (\textbf{u}T) \cdot \mathbf{\hat{j}}\ \rm{d}x = \int_{x} vT \ \rm{d}x,
\end{equation}

\noindent and the RHS becomes

\begin{equation}\label{eq:appB_4}
    \frac{1}{\sqrt{RaPr}} \int_{x} \nabla T  \cdot \mathbf{\hat{j}}\ \rm{d}x = \frac{1}{\sqrt{RaPr}} \int_{x} \frac{\partial T}{\partial y}\  \rm{d}x.
\end{equation}

Note that the integrands in Equations~(\ref{eq:appB_3}) and (\ref{eq:appB_4}) are evaluated at $y=\pm 1$, and the limits of integration in $x$ are from 0 to $2\pi$. The sum of Equations~(\ref{eq:appB_3}) and (\ref{eq:appB_4}) equate to the sum of the convective and conductive surface-integrated heat fluxes:

\begin{equation}\label{eq:appB_5}
    q(t)=q_{\rm{conv}}(t)+q_{\rm{cond}}(t)=\int_{0}^{2\pi} \left ( vT  - \frac{1}{\sqrt{RaPr}} \frac{\partial T}{\partial y} \right ) \ \rm{d}x.
\end{equation}

Multiplying Eq.~\ref{eq:appB_5} with $1/L=1/2\pi$, the resulting terms are then averages of the integrands over $x$ at $y=-1$ and $1$. We represent averages using the angle bracket notation as: 

\begin{equation}\label{eq:appB_6}
    q(t)=\langle vT \rangle_x  - \frac{1}{\sqrt{RaPr}} \frac{\partial \langle T \rangle_x}{\partial y},
\end{equation}

\noindent where we used the Leibnitz integral rule for the last term to interchange the integral and derivative operators. Equation~\ref{eq:q} is then obtained by performing a temporal average of Eq. \ref{eq:appB_6}.

\section*{Appendix C: Extra Materials}\label{AppB}

Extra materials are provided as a series of videos on YouTube, which illustrate:

\begin{itemize}

\item The development and settling of the baseline flow (Fig.~\ref{fig:3_baseline_fields}): see \url{https://www.youtube.com/watch?v=aSKZ1qNgMWk}

\item The RBC undergoing control by the SARL controller, starting from the baseline flow, which illustrates that SARL fails to learn a topology-changing control strategy (Fig.~\ref{fig:4_vanillaDRLvsMARL}, SARL episode 350): see \url{https://www.youtube.com/watch?v=N9wTi4K2uJY}

\item The RBC undergoing successful control by the MARL controller, starting from the baseline flow, which illustrates the bubble coalescence phase and the large bubble phase (Fig.~\ref{fig:6_mechanism}): see \url{https://www.youtube.com/watch?v=_HYmWee3P0A}

\item The modified flow configuration corresponding to a single stable RBC cell, obtained by re-applying a uniform bottom wall temperature at the end of the MARL control sequence (Fig.~\ref{fig:7_removeControl_field}): see \url{https://www.youtube.com/watch?v=MEGl2IiybSg}

\end{itemize}





\end{document}